\documentclass[final,3p,times,sort&compress]{elsarticle}

\usepackage{amssymb}
\usepackage{amsthm}
\usepackage{natbib}

\usepackage{multirow}
\usepackage{amsmath}

\begin{document}

\begin{frontmatter}



\title{Ground-state and magnetocaloric properties of a coupled spin-electron double-tetrahedral chain (exact study at the half filling)}


\author[label1]{Lucia G\'alisov\'a\corref{cor1}}
\cortext[cor1]{Corresponding author}
\ead{galisova.lucia@gmail.com}
\author[label3]{Dorota Jakubczyk}

\address[label1]{Department of Applied Mathematics and Informatics,
					 Faculty of Mechanical Engineering, Technical University,
					 Letn\'a 9, 042 00 Ko\v{s}ice, Slovakia}
\address[label3]{Department of Physics and Medical Engineering,
					 Rzesz\'ow University of Technology,
					 Al. Powsta\'nc\'ow Warszawy~6, 35-959 Rzesz\'ow, Poland}

\begin{abstract}
Ground-state and magnetocaloric properties of a double-tetrahedral chain, in which nodal lattice sites occupied by the localized Ising spins regularly alternate with triangular clusters half filled with mobile electrons, are exactly investigated by using the transfer-matrix method in combination with the construction of the $N$th tensor power of the discrete Fourier transformation. It is shown that the ground state of the model is formed by two non-chiral phases with the zero residual entropy and two chiral phases with the finite residual entropy $S = Nk_{\rm B}\ln 2$. Depending on the character of the exchange interaction between the localized Ising spins and mobile electrons, one or three magnetization plateaus can be observed in the magnetization process. Their heights basically depend on the values of Land\'e $g$-factors of the Ising spins and mobile electrons. It is also evidenced that the system exhibits both the conventional and inverse magnetocaloric effect depending on values of the applied magnetic field and temperature.
\end{abstract}


\end{frontmatter}


\section{Introduction}
\label{intro}

The magnetocaloric effect (MCE), which is defined as the temperature change (i.e., as the cooling or heating) of a magnetic system due to the application of an external magnetic field, has a long history in cooling applications at various temperature regimes~\cite{War81}.
Since the first successful experiment of  the  adiabatic  demagnetization  performed  in 1933~\cite{Gia33}, the MCE represents the standard technique for achieving the extremely low temperatures~\cite{Str07}. In this regard, theoretical predictions and descriptions of materials showing an enhanced MCE create real opportunities for the effective selection of the construction for working magnetic-refrigeration devices. Of particular interest is the investigation of the MCE in various one-dimensional (1D) quantum spin models~\cite{Zhy04, Der06, Hon09, Can09, Tri10, Lan10, Hon11, Top12, Ver13, Kas13, Gal14, Str14a, Zar15, Gal15a} and several coupled spin-electron systems~\cite{Per09, Gal15b, Gal15c, Gal15d}. The reason lies in a possibility of obtaining exact analytical or numerical results as well as in a potential use of these models for the explanation of MCE data measured for real magnetic compounds. In particular, 1D models may give correct quantitative description of many real magnetic compounds when appropriate scaling of material parameters are taken into account~\cite{Hon11, Str05, Str12, Lis14, Bel14}.

In general, the MCE is characterized by the isothermal entropy change ($\Delta S_T$) and/or by the adiabatic temperature change~($\Delta T_{ad}$) upon the magnetic field variation. Depending on sings of these magnetocaloric potentials, the MCE can be either conventional ($\Delta S_T<0$, $\Delta T_{ad}>0$) or inverse ($\Delta S_T>0$, $\Delta T_{ad}<0$). In the former case the system cools down when the magnetic field is removed adiabatically, while in the latter case it heats up. Whether the conventional MCE or the inverse MCE is present basically depends on the particular magnetic arrangement in the system~\cite{Oli10, Tam14a, Tam14b}. Namely, the former phenomenon can be observed in regular ferromagnets or paramagnets, while the latter one can be detected in ferrimagnetic and antiferromagnetic materials. Moreover, coexistence of the above phenomena is also possible. Both conventional and inverse MCE can be found in magnetic systems with rich structure of the ground-state phase diagram, in particular, in various 1D models~\cite{Zhy04, Der06, Hon09, Can09, Tri10, Lan10, Hon11, Top12, Ver13, Kas13, Gal14, Str14a, Zar15, Gal15a, Per09, Gal15b, Gal15c, Gal15d}, some multilayers~\cite{Sza14} or even some finite structures~\cite{Str14b,Str15a,Zuk15}.

In the present paper, we will consider a double-tetrahedral chain, in which nodal lattice sites occupied by the localized Ising spins regularly alternate with triangular clusters with the dynamics described by the Hubbard model. Note that this 1D spin-electron model can be rigorously solved by two distinct analytical approaches regardless of a number of mobile electrons in triangular clusters. The first approach is the standard transfer-matrix technique~\cite{Kra44, Bax82, Str15b}, which is rather straightforward but applicable only to 1D systems. The second one lies in a combination of the generalized decoration-iteration mapping transformation~\cite{Fis59, Syo72, Roj09, Str10} with the well known analytical results for the partition function of the spin-$1/2$ Ising chain in a presence of the longitudinal magnetic field~\cite{Kra44, Bax82, Str15b}. As has been shown in our recent works~\cite{Gal15b, Gal15c, Gal15d}, the coupled spin-electron double-tetrahedral chain provides an excellent prototype model with a rather complex ground state, which allows a rigorous theoretical investigation of the MCE in a vicinity of the first-order phase transitions at non-zero magnetic fields. Last but not least, it is valuable to mention the copper-based polymeric chain Cu$_3$Mo$_2$O$_9$, which represents a possible experimental realization of the double-tetrahedral chain structure~\cite{Has08, Kur10, Kur11, Mat12, Kur14}.

The paper is organized as follows. In the following two Sections~\ref{sec:2} and~\ref{sec:3}, the model under investigation is defined and the corresponding Hilbert space is organized for an exact analytical diagonalization of the block Hamiltonian. Subsequently, a particular ground-state analysis of the model is realized by using a complete set of eigenvalues of the block Hamiltonian. Sections~\ref{sec:4} and~\ref{sec:5} deal with the method used for rigorous analytical solution of the partition function and the most interesting numerical results for the magnetization, entropy and magnetocaloric properties of the model. Finally, the most significant findings are briefly summarized in Section~\ref{sec:5}.

\begin{figure}[t!]
\vspace{-3mm}
\centering
  \includegraphics[width=0.5\textwidth]{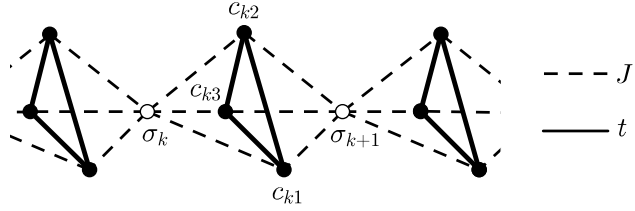}
\caption{A part of the spin-electron system on a double-tetrahedral chain. Empty circles denote nodal lattice sites occupied by the localized Ising spins, while the full circles forming triangular clusters are available to mobile electrons.}
\label{fig:1}
\end{figure}

\section{Spin-electron double-tetrahedral chain}
\label{sec:2}

Let us consider a magnetic system on a double-tetrahedral chain composed of $N$ nodal lattice sites occupied by localized Ising spins and $N$ triangular clusters available to three mobile electrons. The magnetic structure of the considered 1D model is schematically illustrated in Fig.~\ref{fig:1}. Assuming interactions between the nearest neighboring lattice sites, the on-site Coulomb repulsion $U>0$ between two mobile electrons of opposite spins at the same lattice site and the effect of a longitudinal magnetic field $B$ on magnetic particles, the total Hamiltonian of the model reads
\begin{eqnarray}
\label{eq:H_tot}
H\!\!\!&=&\!\!\!
-\,t\sum_{\langle i, j\rangle}\,\sum_{s\in\{\uparrow,\downarrow\}}
\left(c_{i,s}^{\dagger}c_{j,s}+\,c_{j,s}^{\dagger}c_{i,s}\right) + \frac{J}{2}\sum_{\langle j,\, k\rangle}\left(n_{j,\uparrow}-n_{j,\downarrow}\right)\sigma_{k}^z + U\sum_{j}n_{j,\uparrow}n_{j,\downarrow} \nonumber \\
        & &
\!\!\!-\, \frac{g_{e}\mu_{\rm B} B}{2}\sum_{j}\left(n_{j,\uparrow}-n_{j,\downarrow}\right) - g_{I}\mu_{\rm B} B\sum_{k}\sigma_{k}^z.
\end{eqnarray}
Above, the summation $\langle i,\! j\rangle$ runs over lattice sites forming triangular clusters, while the summation $\langle j, k\rangle$ runs over lattice sites of triangular clusters and the nearest-neighboring nodal lattice sites. The operators $c_{i(j),s}^{\dagger}$, $c_{i(j),s}$ represent usual fermionic creation and annihilation operators for mobile electrons occupying the $i(j)$th lattice site with the spin $s\in \{\uparrow,\downarrow\}$, respectively, $n_{j,s}=c_{j,s}^{\dagger}c_{j,s}$ is the number operator of the mobile electron at the $j$th lattice site and $\sigma_{k}^z$ labels the Ising spin localized at the $k$th nodal lattice site. The hopping parameter $t>0$ takes into account the kinetic energy of mobile electrons delocalized over triangular clusters and $J$ stands for the Ising-type coupling between mobile electrons and their nearest Ising neighbors. Finally, the last two terms in~(\ref{eq:H_tot}) represent the Zeeman's energies of the mobile electrons and the localized Ising spins, respectively, in a presence of the applied magnetic field $B$. The quantity $\mu_{\rm B}$ is a Bohr magneton and $g_{e}$, $g_{I}$ denote $g$-factors of the mobile electrons and the localized Ising spins, respectively.

For further calculations, it is advisable to think of the considered model as a system of $N$ interacting double-tetrahedrons whose common vertices are occupied by the localized Ising spins, while others are available for mobile electrons. In this regard, the total Hamiltonian~(\ref{eq:H_tot}) can be written as a sum of $N$ block Hamiltonians  $H=\sum_{k=1}^{N}H_{k}$, where each block Hamiltonian $H_{k}$ involves all the interaction terms associated with triangular clusters available to mobile electrons from $k$th double-tetrahedron (see Fig.~\ref{fig:1})
\begin{eqnarray}
\label{eq:H_k}
H_{k}\!\!\!&=&\!\!\!-\,t\sum_{j\in \tilde{3}}\sum_{s\in\{\uparrow,\downarrow\}}\left(c_{kj,s}^{\dagger}c_{k(j+1)_{\rm mod\,3},s}+c_{k(j+1)_{\rm mod\,3},s}^{\dagger}c_{kj,s}\right)  + U\sum_{j\in \tilde{3}}n_{kj,\uparrow}n_{kj,\downarrow}
- \frac{h_{e}}{2}\sum_{j\in \tilde{3}}\left(n_{kj,\uparrow}-n_{kj,\downarrow}\right)
- h_{I}.
\end{eqnarray}
Above, $\tilde{3}=\{1,2,3\}$ represents a set of lattice sites forming triangular clusters and $h_{e}=-J\!\left(\sigma_{k}^z+\sigma_{k+1}^z\right) + g_{e}\mu_{\rm B} B$, $h_{I}=g_{I}\mu_{\rm B} B\!\left(\sigma_{k}^z+\sigma_{k+1}^z\right)\!/2$ involve all eigenvalues of the Ising spins localized at $k$th and $(k+1)$th nodal lattice sites. From the physical point of view, the former term $h_{e}$ represents the 'effective' field generated by a couple of the Ising spins at $k$th and $(k+1)$th lattice sites, as well as by the external magnetic field acting on mobile electrons. Finally, we have introduced the modulo 3 operation into Eq.~(\ref{eq:H_k}) in order to ensure the periodic boundary condition $c_{k4,s}^{\dagger}=c_{k1,s}^{\dagger}$ ($c_{k4,s}=c_{k1,s}$) for the three-site electron subsystem.

The total Hamiltonian~(\ref{eq:H_tot}) of the model and also the block Hamiltonian~(\ref{eq:H_k}) remain invariant under the action of the cyclic translational operator $c_3 c_{kj,s}^{\dagger}\rightarrow c_{kj+1,s}^{\dagger}$ ($c_3 c_{kj,s}\rightarrow c_{kj+1,s}$) of the cyclic symmetry group $C_3$ as well as under the translation $T\!\!: \sigma_{k}^{z}\rightarrow \sigma_{l}^{z}$. Consequently, the block Hamiltonian~(\ref{eq:H_k}) satisfies the following commutation relations:
\begin{eqnarray}
\label{eq:com_relations1}
\left[H_{k},c_3\right]=\left[H_{k},H_{l}\right]=0.
\end{eqnarray}
Furthermore, each double-tetrahedron of the system shows a conservation of the total number of electrons with the spin up ($n_{k,\uparrow}=\sum_{j\in \tilde{3}}n_{kj,\uparrow}$), the total number of electrons with the spin down ($n_{k,\downarrow}=\sum_{j\in \tilde{3}}n_{kj,\downarrow}$) and the total spin in the $z$th direction $S_{\!k}^{\!z} = \sum_{j\in \tilde{3}}(n_{kj,\uparrow}-n_{kj,\downarrow})/2=(n_{k,\uparrow}-n_{k,\downarrow})/2$ per $k$th triangular cluster. This implies another commutation relations for the  Hamiltonian~(\ref{eq:H_k}):
\begin{eqnarray}
\label{eq:com_relations2}
\left[H_{k},n_{k,\uparrow}\right]=\left[H_{k},n_{k,\downarrow}\right]=\left[H_{k},S_{\!k}^{\!z}\right]=0.
\end{eqnarray}

It is worth noting that the investigated spin-electron chain features further symmetries, except for the aforementioned ones~\cite{Ess05}. However, the relations~(\ref{eq:com_relations1}), (\ref{eq:com_relations2}) are of fundamental importance for an exact analytical diagonalization of the block Hamiltonian~(\ref{eq:H_k}) and subsequent exact analytical evaluation of the partition function as well as all essential thermodynamic quantities.

\section{Ground state}
\label{sec:3}

To determine the ground state of the investigated 1D spin-electron model, it is sufficient to find the lowest-energy eigenstate of the block Hamiltonian~(\ref{eq:H_k}), which can be simply extended to the whole double-tetrahedral chain due to the commuting character of different block Hamiltonians. The lowest-energy eigenstate problem involves the finding of all eigenvalues of the block Hamiltonian~(\ref{eq:H_k}). By assuming the half filling (i.e., three electrons delocalized over each triangular cluster), the relevant calculation can be performed in a matrix representation of the three-site Hilbert subspace ${\cal H}_{ke}$ spanned over all available electron states at $k$th triangular cluster
\begin{eqnarray}
\label{eq:Hilb_e1}
{\cal H}_{ke}= lc_\mathbb{C}\left\{|f_{k\!j}\rangle = c_{k\!j,s_{j}}^{\dagger}c_{kn,s_{n}}^{\dagger}c_{k\!p,s_{p}}^{\dagger}|0\rangle\,\big|\,  j, n, p\in \tilde{3}, s_j, s_n, s_p\in\{\uparrow,\downarrow\}\right\}, \qquad {\rm dim}\, {\cal H}_{ke} = 20.
\end{eqnarray}
In above, the symbol $lc_\mathbb{C}A$ is a linear closure of a set $A$ of all possible electron configurations over the complex field $\mathbb{C}$ and $|0\rangle$ labels the vacuum state. With regard to a conservation of the total spin operator $S_{\!k}^{\!z}$ of mobile electrons at $k$th triangular cluster, the subspace~(\ref{eq:Hilb_e1}) can be decomposed into a tensor sum of four orthogonal sub-spaces ${\cal H}_{ke}^{S_{\!k}^{\!z}}$ with different values of the total spin $S_{\!k}^{\!z}$
\begin{eqnarray}
\label{eq:Hilb_e2}
{\cal H}_{ke}= {\cal H}_{ke}^{-3/2}\oplus{\cal H}_{ke}^{-1/2}\oplus{\cal H}_{ke}^{1/2}\oplus{\cal H}_{ke}^{3/2}.
\end{eqnarray}
The translational invariance of triangular clusters available for mobile electrons allows one to write the relevant sub-spaces ${\cal H}_{ke}^{\,S_k^z}$ as unifications of the orbits ${\cal O}_{|f_{k\!j}^{\,ini}\rangle}$ with well-defined initial electron configuration $|f_{k\!j}^{\,ini}\rangle = c_{k\!j,s_{j}}^{\dagger}c_{kn,s_{n}}^{\dagger}c_{k\!p,s_{p}}^{\dagger}|0\rangle$ that are closed under the the cyclic translation group $C_3$. Each such orbit is formed by electron configuration(s) satisfying the following relations:
\begin{eqnarray}
\label{eq:c3}
c_3|f_{k\!j}^{\,ini}\rangle
\!\!\!&=&\!\!\!c_{k(j+1)_{\rm mod\,3},s_{j}}^{\dagger}c_{k(n+1)_{\rm mod\,3},s_{n}}^{\dagger}c_{k(p+1)_{\rm mod\,3},s_{p}}^{\dagger}|0\rangle,
\nonumber \\
c_3^2|f_{k\!j}^{\,ini}\rangle
\!\!\!&=&\!\!\! c_3\big(c_3|f_{k\!j}^{ini}\rangle\big) = c_{k(j+2)_{\rm mod\,3},s_{j}}^{\dagger}c_{k(n+2)_{\rm mod\,3},s_{n}}^{\dagger}c_{k(p+2)_{\rm mod\,3},s_{p}}^{\dagger}|0\rangle,
\end{eqnarray}
where $c_3\in C_3$ (see Table~\ref{tab:1}). Owing to the validity of the commutation relations~(\ref{eq:com_relations2}), the block Hamiltonian~(\ref{eq:H_k}) can be separately applied to the electron
states of each orthogonal Hilbert sub-space ${\cal H}_{ke}^{\,S_k^z}$. As a result, two $1\times1$ and two $9\times9$ matrices corresponding to the sub-spaces ${\cal H}_{ke}^{\,\mp3/2}$ and ${\cal H}_{ke}^{\,\mp1/2}$, respectively, are generated. The eigenvalues corresponding to ${\cal H}_{ke}^{\,\mp3/2}$ can be found directly (see Table~\ref{tab:1}), while for finding the eigenvalues corresponding to ${\cal H}_{ke}^{\,\mp1/2}$ one should apply the so-called basis of wavelets~\cite{Lul03,Jak09,Jak13} on the orbits ${\cal O}_{|f_{k\!j}^{\,ini}\rangle}\subset{\cal H}_{ke}^{\,\mp1/2}$, as the aftermath of the cyclic translational symmetry of triangular clusters. In our notation, the appropriate amplitude takes the form
\begin{eqnarray}
\label{eq:Fourier}
|\omega, f_{k\!j}^{\,ini}\rangle = \frac{1}{\sqrt{3}}\sum_{j\,\in\,\tilde{3}} \exp\left({\rm i}\omega\!j\right)| f_{k\!j}, f_{k\!j}^{\,ini}\rangle,
\end{eqnarray}
where $| f_{k\!j}, f_{k\!j}^{\,ini}\rangle$ denotes the $j$th electron configuration of the orbit ${\cal O}_{|f_{k\!j}^{\,ini}\rangle}$ and the discrete quasi-momentum $\omega = 2\pi b/3$ corresponds to the Brillouin zone $B = \{b = 0, \mp 1\}$. The $9\times9$ matrix sectors corresponding to the sub-spaces ${\cal H}_{ke}^{\,\mp1/2}$ take quasi-diagonal forms with three disjoint sub-sectors corresponding to each orbits ${\cal O}_{|f_{k\!j}^{\,ini}\rangle}\subset{\cal H}_{ke}^{\,\mp1/2}$ in the basis of wavelets~(\ref{eq:Fourier}). After a direct diagonalization of relevant sub-sectors, the remaining eighteen eigenvalues of the block Hamiltonian~(\ref{eq:H_k}), corresponding to the relevant orbits ${\cal O}_{|f_{k\!j}^{\,ini}\rangle}\subset{\cal H}_{ke}^{\,\mp1/2}$, can be obtained (see Table~\ref{tab:1}).

At this stage, a comprehensive analysis of all possible ground states of the considered model can be carried out by inspection of the complete spectrum of eigenvalues listed in Table~\ref{tab:1} for all available configurations of the nodal Ising spins $\sigma_{k}$ and $\sigma_{k+1}$ involved therein. It is worthwhile to remark, however, that there are some fundamental differences between magnetic behavior of the model when distinct nature of the Ising exchange interaction $J$ is considered (see our recent works~\cite{Gal15b,Gal15c,Gal15d}). The ground-state analysis will be therefore divided into the case of the ferromagnetic interaction $J<0$ and the case of the antiferromagnetic interaction $J>0$ between the localized Ising spins and mobile electrons.

\begin{figure}[h!]
\centering
\vspace{-0.1cm}
  \includegraphics[width=1.0\textwidth]{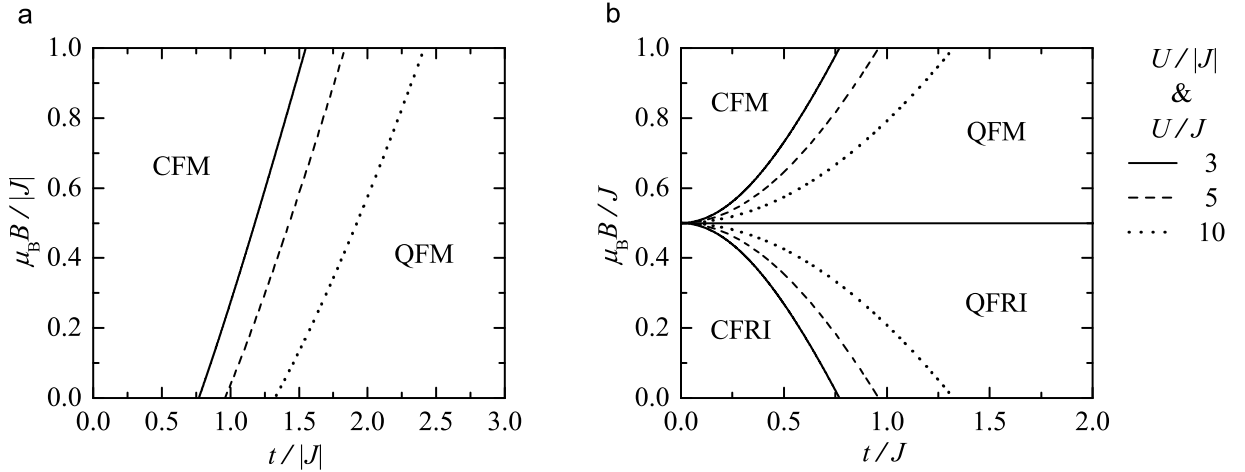}
  \vspace{-0.75cm}
\caption{Ground-state phase diagrams in the $t - B$ plane for the spin-electron double-tetrahedral chain with (a)~the ferromagnetic Ising interaction $J<0$ and (b)~the antiferromagnetic Ising interaction $J>0$, by assuming the Land\'e $g$-factors $g_e = 2$, $g_I \geq 6$ and three fixed values of the Coulomb term~$U/|J| = U/J = 3, 5, 10$.}
\label{fig:2}
\vspace{0.0cm}
\end{figure}

The ground-state phase diagram of the model with $J<0$ is shown in Fig.~\ref{fig:2}a. As one can see, the ground state of this model consists of the classical ferromagnetic (CFM) phase and the quantum ferromagnetic (QFM) phase due to a mutual interplay between model parameters and the applied magnetic field. The boundary, which represents the first-order transition between these two phases, is given by the condition
\begin{eqnarray}
\label{eq:Bc_ferro}
\mu_{\rm B}B_{c1} = \frac{1}{g_e}\left\{J - \frac{2U}{3} + \frac{2}{3}\sqrt{U^2 + 27t^2}\cos\left[\frac{1}{3}\arctan\left(\frac{9t}{U^3}\sqrt{U^4 + 27U^2t^2 + 243t^4}\right)\right] \right\}\!.
\end{eqnarray}
The spin-electron arrangements peculiar to the CFM and QFM phases are unambiguously determined by the following eigenvectors and the corresponding energies:
\begin{eqnarray}
\label{eq:CFM}
|{\rm CFM}\rangle\!\!\!&=&\!\!\!
\begin{cases}
\lefteqn{{}\prod\limits_{k=1}^N
    \Bigg\{\begin{array}{l}
        |\!\downarrow\rangle_{\sigma_k}\!\otimes|\Psi^{\,-3/2}\rangle_{k} \\
       |\!\uparrow\rangle_{\sigma_k}\!\otimes|\Psi^{\,3/2}\rangle_{k}
           \end{array}\quad \text{for $B=0$},
         \nonumber
         }
\\[2mm]                          \lefteqn{{}\prod\limits_{k=1}^N|\!\uparrow\rangle_{\sigma_k}\!\otimes|\Psi^{\,3/2}\rangle_{k}
        \hspace{11mm} \text{for $B\neq0$},
        }
\end{cases}
\\
E_{\rm CFM}\!\!\!&=&\!\!\!\frac{N}{2}\bigg[3J - \mu_{\rm B}B(g_I+3g_e)\bigg];
\hspace{9.5cm}
\end{eqnarray}
\begin{eqnarray}
\label{eq:QFM}
|{\rm QFM}\rangle\!\!\!&=&\!\!\!
\begin{cases}
\lefteqn{{}\prod\limits_{k=1}^N
    \Bigg\{\begin{array}{l}
        |\!\downarrow\rangle_{\sigma_k}\!\otimes|\Psi^{\,-1/2}\rangle_{k} \\
       |\!\uparrow\rangle_{\sigma_k}\!\otimes|\Psi^{\,1/2}\rangle_{k}
           \end{array}\quad \text{for $B=0$},
         \nonumber
         }
\\[2mm]                          \lefteqn{{}\prod\limits_{k=1}^N|\!\uparrow\rangle_{\sigma_k}\!\otimes|\Psi^{\,1/2}\rangle_{k}
        \hspace{11mm} \text{for $B\neq0$},
        }
\end{cases}
\\
E_{{\rm QFM}}\!\!\!&=&\!\!\!\frac{N}{6}\left\{3J + 4U - 4\!\sqrt{U^2 + 27t^2}\cos\left[\frac{1}{3}\arctan\left(\frac{9t}{U^3}\sqrt{U^4 + 27U^2t^2 + 243t^4}\right)\right] - 3\mu_{\rm B}B(g_I+g_e)\right\}.
\end{eqnarray}
In above, the product $\prod_{k=1}^N$ runs over all primitive unit cells, the state vector $|\!\uparrow\rangle_{\sigma_k}$ $(|\!\downarrow\rangle_{\sigma_k})$ determines up (down) state of the Ising spin $\sigma_k^z = 1/2$ ($\sigma_k^z = -1/2$) localized at $k$th nodal lattice site. The state vectors $|\Psi^{\,\mp3/2}\rangle_{k}$, $|\Psi^{\,\mp1/2}\rangle_{k}$ label the eigenstates of the mobile electrons from $k$th triangular cluster with the fixed value of the total spin $S_k^z=\pm3/2, \pm1/2$, respectively:
\begin{table}[b!]
\caption{Decomposition of three electrons from $k$th triangular cluster into the orbits ${\cal O}_{|f_{k\!j}^{\,ini}\rangle}$ of the cyclic symmetry group~$C_{3}$ and the corresponding eigenvalues $E_{kl}$ $(l = 1,2, 3,\ldots, 20)$ of the block Hamiltonian~(\ref{eq:H_k}). The parameter $\phi$ used in the notation of $E_{kl}$ depends on the terms $t$ and $U$ through the relation $\tan \left(3\phi\right) = 9t\sqrt{U^4+27U^2t^2+243t^4}/U^3$.}
\vspace{5mm}
\label{tab:1}
\centering
\begin{tabular}{rccl}
\hline\noalign{\smallskip}
$S_k^z$ & $|f_{k\!j}^{\,ini}\rangle$ & ${\cal O}_{|f_{k\!j}^{\,ini}\rangle}$ & $E_{kl}$\\
\noalign{\smallskip}
\hline\hline\noalign{\smallskip}
\multirow{1}{*}{$-\frac{3}{2}$} &
        $c_{k1,\downarrow}^{\dagger}c_{k2,\downarrow}^{\dagger}c_{k3,\downarrow}^{\dagger}|0\rangle$ &
        $c_{k1,\downarrow}^{\dagger}c_{k2,\downarrow}^{\dagger}c_{k3,\downarrow}^{\dagger}|0\rangle$ &
        $-h_I+\frac{3h_e}{2}$
        \\
   \noalign{\smallskip}\hline\noalign{\smallskip}
\multirow{1}{*}{$-\frac{1}{2}$} &
        $c_{k1,\downarrow}^{\dagger}c_{k2,\downarrow}^{\dagger}c_{k3,\uparrow}^{\dagger}|0\rangle$ &
        $c_{k1,\downarrow}^{\dagger}c_{k2,\downarrow}^{\dagger}c_{k3,\uparrow}^{\dagger}|0\rangle$ &
        $-h_I+\frac{h_e}{2}$
        \\
   & &
        $c_{k2,\downarrow}^{\dagger}c_{k3,\downarrow}^{\dagger}c_{k1,\uparrow}^{\dagger}|0\rangle$ &
        $-h_I+\frac{h_e}{2} + U$
        \\
   & &
        $c_{k3,\downarrow}^{\dagger}c_{k1,\downarrow}^{\dagger}c_{k2,\uparrow}^{\dagger}|0\rangle$ &
        $-h_I+\frac{h_e}{2} + U$
        \\	
    \noalign{\smallskip}\cline{2-4}\noalign{\smallskip}
   & $c_{k2,\uparrow}^{\dagger}c_{k2,\downarrow}^{\dagger}c_{k3,\downarrow}^{\dagger}|0\rangle$ &
     $c_{k2,\uparrow}^{\dagger}c_{k2,\downarrow}^{\dagger}c_{k3,\downarrow}^{\dagger}|0\rangle$ &
     $-h_I+\frac{h_e}{2}+\frac{2U}{3}-\frac{2}{3}\!\sqrt{U^2+27t^2}\cos\left(\phi\right)$
        \\
   & &
        $c_{k3,\uparrow}^{\dagger}c_{k3,\downarrow}^{\dagger}c_{k1,\downarrow}^{\dagger}|0\rangle$ &
     $-h_I+\frac{h_e}{2}+\frac{2U}{3}-\frac{2}{3}\!\sqrt{U^2+27t^2}\cos\left(\phi + \frac{2\pi}{3}\right)$
        \\
   & &
        $c_{k1,\uparrow}^{\dagger}c_{k1,\downarrow}^{\dagger}c_{k2,\downarrow}^{\dagger}|0\rangle$ &
     $-h_I+\frac{h_e}{2}+\frac{2U}{3}-\frac{2}{3}\!\sqrt{U^2+27t^2}\cos\left(\phi + \frac{4\pi}{3}\right)$
        \\
     \noalign{\smallskip}\cline{2-4}\noalign{\smallskip}
	&   $c_{k3,\downarrow}^{\dagger}c_{k1,\uparrow}^{\dagger}c_{k1,\downarrow}^{\dagger}|0\rangle$ &
        $c_{k3,\downarrow}^{\dagger}c_{k1,\uparrow}^{\dagger}c_{k1,\downarrow}^{\dagger}|0\rangle$ &
     $-h_I+\frac{h_e}{2}+\frac{2U}{3}-\frac{2}{3}\!\sqrt{U^2+27t^2}\cos\left(\phi\right)$
        \\
   & &
        $c_{k1,\downarrow}^{\dagger}c_{k2,\uparrow}^{\dagger}c_{k2,\downarrow}^{\dagger}|0\rangle$ &
     $-h_I+\frac{h_e}{2}+\frac{2U}{3}-\frac{2}{3}\!\sqrt{U^2+27t^2}\cos\left(\phi + \frac{2\pi}{3}\right)$
        \\
   & &
        $c_{k2,\downarrow}^{\dagger}c_{k3,\uparrow}^{\dagger}c_{k3,\downarrow}^{\dagger}|0\rangle$ &
     $-h_I+\frac{h_e}{2}+\frac{2U}{3}-\frac{2}{3}\!\sqrt{U^2+27t^2}\cos\left(\phi + \frac{4\pi}{3}\right)$
        \\
   \noalign{\smallskip}\hline\noalign{\smallskip}
 \multirow{1}{*}{$\frac{1}{2}$} &
        $c_{k1,\uparrow}^{\dagger}c_{k2,\uparrow}^{\dagger}c_{k3,\downarrow}^{\dagger}|0\rangle$ &
        $c_{k1,\uparrow}^{\dagger}c_{k2,\uparrow}^{\dagger}c_{k3,\downarrow}^{\dagger}|0\rangle$ &
        $-h_I-\frac{h_e}{2}$
        \\
    & &
        $c_{k2,\uparrow}^{\dagger}c_{k3,\uparrow}^{\dagger}c_{k1,\downarrow}^{\dagger}|0\rangle$ &
        $-h_I-\frac{h_e}{2} + U$
        \\
    & &
        $c_{k3,\uparrow}^{\dagger}c_{k1,\uparrow}^{\dagger}c_{k2,\downarrow}^{\dagger}|0\rangle$ &
        $-h_I-\frac{h_e}{2} + U$
        \\
	\noalign{\smallskip}\cline{2-4}\noalign{\smallskip}
	 &  $c_{k2,\uparrow}^{\dagger}c_{k2,\downarrow}^{\dagger}c_{k3,\uparrow}^{\dagger}|0\rangle$ &
        $c_{k2,\uparrow}^{\dagger}c_{k2,\downarrow}^{\dagger}c_{k3,\uparrow}^{\dagger}|0\rangle$ &
     $-h_I-\frac{h_e}{2}+\frac{2U}{3}-\frac{2}{3}\!\sqrt{U^2+27t^2}\cos\left(\phi\right)$
        \\
    & &
        $c_{k3,\uparrow}^{\dagger}c_{k3,\downarrow}^{\dagger}c_{k1,\uparrow}^{\dagger}|0\rangle$ &
     $-h_I-\frac{h_e}{2}+\frac{2U}{3}-\frac{2}{3}\!\sqrt{U^2+27t^2}\cos\left(\phi+ \frac{2\pi}{3}\right)$
        \\
    & &
        $c_{k1,\uparrow}^{\dagger}c_{k1,\downarrow}^{\dagger}c_{k2,\uparrow}^{\dagger}|0\rangle$ &
     $-h_I-\frac{h_e}{2}+\frac{2U}{3}-\frac{2}{3}\!\sqrt{U^2+27t^2}\cos\left(\phi+ \frac{4\pi}{3}\right)$
        \\
	\noalign{\smallskip}\cline{2-4}\noalign{\smallskip}
    &   $c_{k3,\uparrow}^{\dagger}c_{k1,\uparrow}^{\dagger}c_{k1,\downarrow}^{\dagger}|0\rangle$ &
        $c_{k3,\uparrow}^{\dagger}c_{k1,\uparrow}^{\dagger}c_{k1,\downarrow}^{\dagger}|0\rangle$ &
     $-h_I-\frac{h_e}{2}+\frac{2U}{3}-\frac{2}{3}\!\sqrt{U^2+27t^2}\cos\left(\phi\right)$
        \\
    & &
        $c_{k1,\uparrow}^{\dagger}c_{k2,\uparrow}^{\dagger}c_{k2,\downarrow}^{\dagger}|0\rangle$ &
     $-h_I-\frac{h_e}{2}+\frac{2U}{3}-\frac{2}{3}\!\sqrt{U^2+27t^2}\cos\left(\phi+ \frac{2\pi}{3}\right)$
        \\
    & &
        $c_{k2,\uparrow}^{\dagger}c_{k3,\uparrow}^{\dagger}c_{k3,\downarrow}^{\dagger}|0\rangle$ &
     $-h_I-\frac{h_e}{2}+\frac{2U}{3}-\frac{2}{3}\!\sqrt{U^2+27t^2}\cos\left(\phi+ \frac{4\pi}{3}\right)$
        \\
    \noalign{\smallskip}\hline\noalign{\smallskip}
\multirow{1}{*}{$\frac{3}{2}$} &
        $c_{k1,\uparrow}^{\dagger}c_{k2,\uparrow}^{\dagger}c_{k3,\uparrow}^{\dagger}|0\rangle$ &
        $c_{k1,\uparrow}^{\dagger}c_{k2,\uparrow}^{\dagger}c_{k3,\uparrow}^{\dagger}|0\rangle$ &
        $-h_I-\frac{3h_e}{2}$
        \\
\noalign{\smallskip}\hline
\end{tabular}
\end{table}
\begin{eqnarray}
\label{eq:eigenstatesPSI-3/2}
|\Psi^{\,-3/2}\rangle_{k}\!\!\!&=&\!\!\!
c_{k1,\downarrow}^{\dagger}c_{k2,\downarrow}^{\dagger}c_{k3,\downarrow}^{\dagger}|0\rangle,
\\
\label{eq:eigenstatesPSI3/2}
|\Psi^{\,3/2}\rangle_{k}\!\!\!&=&\!\!\!
c_{k1,\uparrow}^{\dagger}c_{k2,\uparrow}^{\dagger}c_{k3,\uparrow}^{\dagger}|0\rangle,
\\
\label{eq:eigenstatesPSI-1/2}
|\Psi^{\,-1/2}\rangle_{k}\!\!\!&=&\!\!\!
\begin{cases}
\,
\Big[{\cal A} \left(c_{k1,\downarrow}^{\dagger}c_{k2,\downarrow}^{\dagger}c_{k3,\uparrow}^{\dagger}
             +{\rm e}^{\frac{2\pi\,{\rm i}}{3}}c_{k3,\downarrow}^{\dagger}c_{k1,\downarrow}^{\dagger}c_{k2,\uparrow}^{\dagger}
            +{\rm e}^{\frac{4\pi\,{\rm i}}{3}}c_{k2,\downarrow}^{\dagger}c_{k3,\downarrow}^{\dagger}c_{k1,\uparrow}^{\dagger}\right)
    +{\cal B} \left(c_{k1,\uparrow}^{\dagger}c_{k1,\downarrow}^{\dagger}c_{k2,\downarrow}^{\dagger}
            -c_{k3,\downarrow}^{\dagger}c_{k1,\uparrow}^{\dagger}c_{k1,\downarrow}^{\dagger}     \right)
\\[1mm]
	\hspace{4mm}+\,{\cal C} \left(c_{k2,\downarrow}^{\dagger}c_{k3,\uparrow}^{\dagger}c_{k3,\downarrow}^{\dagger}-
      c_{k2,\uparrow}^{\dagger}c_{k2,\downarrow}^{\dagger}c_{k3,\downarrow}^{\dagger}
            \right)
	+ {\cal D} \left(c_{k3,\uparrow}^{\dagger}c_{k3,\downarrow}^{\dagger}c_{k1,\downarrow}^{\dagger}
            -c_{k1,\downarrow}^{\dagger}c_{k2,\uparrow}^{\dagger}c_{k2,\downarrow}^{\dagger}     \right)
           \Big]|0\rangle{}
\\[1mm]
\,
\Big[{\cal A} \left(c_{k1,\downarrow}^{\dagger}c_{k2,\downarrow}^{\dagger}c_{k3,\uparrow}^{\dagger}
             +{\rm e}^{\frac{4\pi\,{\rm i}}{3}}c_{k3,\downarrow}^{\dagger}c_{k1,\downarrow}^{\dagger}c_{k2,\uparrow}^{\dagger}
            +{\rm e}^{\frac{2\pi\,{\rm i}}{3}}c_{k2,\downarrow}^{\dagger}c_{k3,\downarrow}^{\dagger}c_{k1,\uparrow}^{\dagger}\right)
    +{\cal B} \left(c_{k1,\uparrow}^{\dagger}c_{k1,\downarrow}^{\dagger}c_{k2,\downarrow}^{\dagger}
            -c_{k3,\downarrow}^{\dagger}c_{k1,\uparrow}^{\dagger}c_{k1,\downarrow}^{\dagger}     \right)
\\[1mm]
	\hspace{4mm}+\,{\cal C} \left(c_{k2,\downarrow}^{\dagger}c_{k3,\uparrow}^{\dagger}c_{k3,\downarrow}^{\dagger}-
      c_{k2,\uparrow}^{\dagger}c_{k2,\downarrow}^{\dagger}c_{k3,\downarrow}^{\dagger}
              \right)
	+ {\cal D} \left(c_{k3,\uparrow}^{\dagger}c_{k3,\downarrow}^{\dagger}c_{k1,\downarrow}^{\dagger}
            -c_{k1,\downarrow}^{\dagger}c_{k2,\uparrow}^{\dagger}c_{k2,\downarrow}^{\dagger}     \right)
           \Big]|0\rangle{},
\end{cases}
\\
\label{eq:eigenstatesPSI1/2}
|\Psi^{\,1/2}\rangle_{k}\!\!\!&=&\!\!\!
\begin{cases}
\,
\Big[{\cal A} \left(c_{k3,\uparrow}^{\dagger}c_{k1,\uparrow}^{\dagger}c_{k2,\downarrow}^{\dagger}
             +{\rm e}^{\frac{2\pi\,{\rm i}}{3}}c_{k1,\uparrow}^{\dagger}c_{k2,\uparrow}^{\dagger}c_{k3,\downarrow}^{\dagger}
            +{\rm e}^{\frac{4\pi\,{\rm i}}{3}}c_{k2,\uparrow}^{\dagger}c_{k3,\uparrow}^{\dagger}c_{k1,\downarrow}^{\dagger}\right)
    +{\cal B} \left(c_{k1,\uparrow}^{\dagger}c_{k1,\downarrow}^{\dagger}c_{k2,\uparrow}^{\dagger}
            - c_{k3,\uparrow}^{\dagger}c_{k1,\uparrow}^{\dagger}c_{k1,\downarrow}^{\dagger} \right){}
\\[1mm]
	\hspace{4mm}+\,{\cal C} \left(c_{k2,\uparrow}^{\dagger}c_{k3,\uparrow}^{\dagger}c_{k3,\downarrow}^{\dagger}-
      c_{k2,\uparrow}^{\dagger}c_{k2,\downarrow}^{\dagger}c_{k3,\uparrow}^{\dagger}\right)
  +{\cal D}\left(c_{k3,\uparrow}^{\dagger}c_{k3,\downarrow}^{\dagger}c_{k1,\uparrow}^{\dagger}
            - c_{k1,\uparrow}^{\dagger}c_{k2,\uparrow}^{\dagger}c_{k2,\downarrow}^{\dagger}\right)
\Big]|0\rangle{}
\\[1mm]
\,
\Big[{\cal A} \left(c_{k3,\uparrow}^{\dagger}c_{k1,\uparrow}^{\dagger}c_{k2,\downarrow}^{\dagger}
             +{\rm e}^{\frac{4\pi\,{\rm i}}{3}}c_{k1,\uparrow}^{\dagger}c_{k2,\uparrow}^{\dagger}c_{k3,\downarrow}^{\dagger}
            +{\rm e}^{\frac{2\pi\,{\rm i}}{3}}c_{k2,\uparrow}^{\dagger}c_{k3,\uparrow}^{\dagger}c_{k1,\downarrow}^{\dagger}\right)
    +{\cal B} \left(c_{k1,\uparrow}^{\dagger}c_{k1,\downarrow}^{\dagger}c_{k2,\uparrow}^{\dagger}
            - c_{k3,\uparrow}^{\dagger}c_{k1,\uparrow}^{\dagger}c_{k1,\downarrow}^{\dagger} \right){}
\\[1mm]
	\hspace{4mm}+\,{\cal C} \left(c_{k2,\uparrow}^{\dagger}c_{k3,\uparrow}^{\dagger}c_{k3,\downarrow}^{\dagger}-
      c_{k2,\uparrow}^{\dagger}c_{k2,\downarrow}^{\dagger}c_{k3,\uparrow}^{\dagger}\right)
  +{\cal D}\left(c_{k3,\uparrow}^{\dagger}c_{k3,\downarrow}^{\dagger}c_{k1,\uparrow}^{\dagger}
            - c_{k1,\uparrow}^{\dagger}c_{k2,\uparrow}^{\dagger}c_{k2,\downarrow}^{\dagger}\right)
\Big]|0\rangle{}.
\end{cases}
\end{eqnarray}
The coefficients ${\cal A}$, ${\cal B}$, ${\cal C}$, ${\cal D}$ emerging in the last two state vectors~(\ref{eq:eigenstatesPSI-1/2}) and (\ref{eq:eigenstatesPSI1/2}) determine a quantum entanglement of the relevant electron states within these eigenstates. However, analytical expressions for the above coefficients are too cumbersome, therefore we do not write them here explicitly. We just note that they are the functions of the model parameters $t$, $U$ and $J$.

As one can see from Eqs. (\ref{eq:CFM}) and (\ref{eq:QFM}), the common feature of CFM and QFM phases is that the localized Ising spins as well as mobile electrons may choose between two energetically equivalent states if no magnetic field is applied on magnetic particles. In this particular case, the Ising spins are free to choose either the down or up state, while the mobile electrons from each triangular cluster choose between two classical ferromagtnetic states with the the total spins $S_k^z = -3/2$ and $3/2$ (in the CFM phase) or between two quantum states corresponding to $S_k^z = -1/2$ and $1/2$ (in the QFM phase), in order to preserve the spontaneous ferromagnetic order with the nearest Ising neighbors. Arbitrary but non-zero magnetic field lifts this two-fold degeneracy, since it tends to align all Ising spins into the external-field direction. Thus, if $B\neq0$, the CFM phase exhibits the unique ferromagnetic arrangement with all Ising spins and mobile electrons fully polarized to the external-field direction. By contrast, the QFM phase remains macroscopically degenerate for any $B\in\left\langle0, B_{c1}\right)$ due to two possible (right- and left-hand side) chiral degrees of freedom of mobile electrons from each triangular cluster (see Eqs.~(\ref{eq:eigenstatesPSI-1/2}) and (\ref{eq:eigenstatesPSI1/2})).
\newpage
It is obvious from the phase diagram shown in Fig.~\ref{fig:2}b that the CFM and QFM phases also appear in the ground state of the antiferromagnetic counterpart of the model when the applied magnetic field $B$ is stronger than the critical value
\begin{eqnarray}
\label{eq:Bc_UFM-FRI2}
\mu_{\rm B}B_{c2} = \frac{J}{g_e}.
\end{eqnarray}
Furthermore, there also arise another two ferrimagnetic phases CFRI and QFRI as an response to the mutual interplay between the model parameters $J>0$, $t$, $U$ and the magnetic field $B<B_{c2}$. The observed CFRI and QFRI phases are unambiguously characterized by the following eigenvectors and energies:
\begin{eqnarray}
\label{eq:CFRI}
|{\rm CFRI}\rangle\!\!\!&=&\!\!\!
\begin{cases}
\lefteqn{{}\prod\limits_{k=1}^N
    \Bigg\{\begin{array}{l}
        |\!\downarrow\rangle_{\sigma_k}\!\otimes|\Psi^{\,3/2}\rangle_{k} \\
       |\!\uparrow\rangle_{\sigma_k}\!\otimes|\Psi^{\,-3/2}\rangle_{k}
           \end{array}\quad \text{for $B=0$ and for $B\neq0$ if $g_I=3g_e$},
         \nonumber
         }
\\[2mm]                          \lefteqn{{}\prod\limits_{k=1}^N|\!\uparrow\rangle_{\sigma_k}\!\otimes|\Psi^{\,-3/2}\rangle_{k}
        \hspace{9mm} \text{for $B\neq0$ if $g_I>3g_e$},
        }
\end{cases}
\\
E_{{\rm CFRI}}\!\!\!&=&\!\!\!-\frac{N}{2}\bigg[3J + \mu_{\rm B}B(g_I-3g_e)\bigg];
\hspace{9.4cm}
\end{eqnarray}
\begin{eqnarray}
\label{eq:FRI2}
|{\rm QFRI}\rangle\!\!\!&=&\!\!\!
\begin{cases}
\lefteqn{{}\prod\limits_{k=1}^N
    \Bigg\{\begin{array}{l}
        |\!\downarrow\rangle_{\sigma_k}\!\otimes|\Psi^{\,1/2}\rangle_{k} \\
       |\!\uparrow\rangle_{\sigma_k}\!\otimes|\Psi^{\,-1/2}\rangle_{k}
           \end{array}\quad \text{for $B=0$},
         \nonumber
         }
\\[2mm]                          \lefteqn{{}\prod\limits_{k=1}^N|\!\uparrow\rangle_{\sigma_k}\!\otimes|\Psi^{\,-1/2}\rangle_{k}
        \hspace{9mm} \text{for $B\neq0$},
        }
\end{cases}
\\
E_{{\rm QFRI}}\!\!\!&=&\!\!\!-\frac{N}{6}\left\{3J - 4U + 4\!\sqrt{U^2 + 27t^2}\cos\left[\frac{1}{3}\arctan\left(\frac{9t}{U^3}\sqrt{U^4 + 27U^2t^2 + 243t^4}\right)\right]+ 3\mu_{\rm B}B(g_I-g_e)\right\}.
\end{eqnarray}
Evidently, the ${\rm CFRI}$ and ${\rm QFRI}$ phases basically differ from the CFM and QFM phases just in a relevant alignment of the localized Ising spins, which are oriented antiparallel (parallel) with respect to the total spin $S_k^z$ of mobile electrons from triangular clusters in ${\rm CFRI}$, ${\rm QFRI}$ (CFM, QFM). Moreover, the former ${\rm CFRI}$ phase is two-fold degenerate not only in the zero-field region, but, surprisingly, also at non-zero magnetic fields if Land\'e $g$-factors of the Ising spins and mobile electrons satisfy the condition $g_I = 3g_e$. This phase becomes a ground state whenever the mutual interplay between the antiferromagnetic Ising interaction $J>0$, the hopping term $t$ and the Coulomb term $U$ overwhelms the effect of the external magnetic field, i.e. whenever the applied field $B$ is weaker than the critical value
\begin{eqnarray}
\label{eq:Bc_FRI1-FRI2}
\mu_{\rm B}B_{c3} = \frac{1}{g_e}\left\{J + \frac{2U}{3} - \frac{2}{3}\sqrt{U^2 + 27t^2}\cos\left[\frac{1}{3}\arctan\left(\frac{9t}{U^3}\sqrt{U^4 + 27U^2t^2 + 243t^4}\right)\right] \right\}\!.
\end{eqnarray}
If the reverse condition $B > B_{c3}$ holds, but the intensity of the applied field is still weaker than the antiferromagnetic coupling $J$, the macroscopically degenerate ${\rm QFRI}$ phase with two possible chiral degrees of freedom of mobile electrons on each triangular cluster is preferred as a ground state due to a prevailing influence of the kinetic term $t$.

\section{Magnetization process and entropy}
\label{sec:4}

A crucial step for the investigation of thermodynamic quantities of the considered spin-electron double-tetrahedral chain lies in exact derivation of the partition function for this model. Having a full spectrum of eigenvalues of the block Hamiltonian~(\ref{eq:H_k}) listed in Table~\ref{tab:1}, the partition function $Z$ of the studied model can be easily found by means of the standard transfer-matrix approach~\cite{Kra44, Bax82, Str15b}:
\begin{eqnarray}
\label{eq:Z}
Z = \sum_{\{\sigma_k\}}\prod_{k = 1}^N{\rm Tr}_k{\rm exp}\left(-\beta H_k\right) = \sum_{\{\sigma_k\}}\prod_{k = 1}^N\sum_{l=1}^{20}{\rm exp}\left(-\beta E_{kl}\right) = \sum_{\{\sigma_k\}}\prod_{k = 1}^N {\rm T}\big(\sigma_k^z, \sigma_{k+1}^z\big) = \sum_{\{\sigma_k\}} {\rm T}^N\big(\sigma_k^z, \sigma_{k+1}^z\big) = \lambda_{+}^N + \lambda_{-}^N ,
\end{eqnarray}
where $\beta = 1/(k_{\rm B}T)$ is the inverse temperature ($k_{\rm B}$  is the Boltzmann's constant and $T$ is the absolute temperature), the symbol $\sum_{\{\sigma_k\}}$ denotes the summation over all possible states of the localized Ising spins and the symbol ${\rm Tr}_k$ labels a trace over degrees of freedom of mobile electrons from the $k$th triangular cluster. The expression ${\rm T}\big(\sigma_k^z, \sigma_{k+1}^z\big)$ can be viewed as the $2\times2$ transfer matrix
\begin{eqnarray}
\label{eq:TransferMatrix}
{\rm T}\big(\sigma_k^z, \sigma_{k+1}^z\big)
=
\left(
\begin{array}{cc}
{\rm T}_{1/2,1/2}  & {\rm T}_{1/2,-1/2}  \\
{\rm T}_{-1/2,1/2}  & {\rm T}_{-1/2,-1/2}  \\
\end{array}
\right),
\end{eqnarray}
whose elements depend just on values of the localized Ising spins $\sigma_k^z$ and $\sigma_{k+1}^z$ according to the formula
\begin{eqnarray}
\label{eq:TMatrixElements}
{\rm T}_{\sigma_k^z,\,\sigma_{k+1}^z} =
\sum_{l=1}^{20}{\rm exp}\left(-\beta E_{kl}\right)=2{\rm exp}\left(\beta h_I\right)
\Bigg\{
\cosh\left(\frac{3\beta h_e}{2}\right) + \cosh\left(\frac{\beta h_e}{2}\right)
+2{\rm exp}\left(-\beta U\right)\cosh\left(\frac{\beta h_e}{2}\right)
\hspace{20mm}
\nonumber\\
 +\,{\rm exp}\left(\frac{-2\beta U}{3}\right)\cosh\left(\frac{\beta h_e}{2}\right)\sum_{q=1}^3{\rm exp}\left(\frac{2\beta}{3}\sqrt{U^2+27t^2}\right)\cos\left[\frac{1}{3}\arctan\left(\frac{9t}{U^3}\sqrt{U^4 + 27U^2t^2 + 243t^4}\right) + \frac{2\pi q}{3}\right]
\Bigg\}.
\end{eqnarray}
Recall, the values of the Ising spins $\sigma_k^z$ and $\sigma_{k+1}^z$ are included in terms $h_I$ and $h_e$ (see the text below Eq.~(\ref{eq:H_k})). The eigenvalues $\lambda_{\pm}$ emerging at the end of the partition function~(\ref{eq:Z}) can be calculated by solving the eigenvalue problem for the transfer matrix~(\ref{eq:TransferMatrix}):
\begin{eqnarray}
\label{eq:TMatrixEigenvalues}
\lambda_{\pm}\!\!\!&=&\!\!\! \frac{1}{2}\left[{\rm T}_{1/2,1/2} + {\rm T}_{-1/2,-1/2} \pm \sqrt{\left({\rm T}_{1/2,1/2}-{\rm T}_{-1/2,-1/2}\right)^2 + 4{\rm T}_{-1/2,1/2}{\rm T}_{1/2,-1/2}}\,\right].
\end{eqnarray}
At this stage, the exact calculation of the partition function of the investigated spin-electron double-tetrahedral chain is formally completed. For calculation of all thermodynamic quantities of the model is now sufficient to substitute the exact result~(\ref{eq:TMatrixEigenvalues}) into the partition function~(\ref{eq:Z}) and subsequently calculate the Gibbs free energy $G$ of the spin-electron model, which depends just on the larger eigenvalue of the transfer matrix~(\ref{eq:TransferMatrix}) in the thermodynamic limit $N\to\infty$:
\begin{eqnarray}
\label{eq:G}
G \!\!\!&=&\!\!\! -k_{\rm B}T\lim_{N\to\infty}\ln Z =
-k_{\rm B}T\lim_{N\to\infty}\ln \left(\lambda_{+}^N + \lambda_{-}^N\right) = -k_{\rm B}TN\ln \lambda_{+}
\nonumber\\
\!\!\!&=&\!\!\!
k_{\rm B}TN\ln2 - k_{\rm B}TN\ln\left[{\rm T}_{1/2,1/2} + {\rm T}_{-1/2,-1/2} \pm \sqrt{\left({\rm T}_{1/2,1/2}-{\rm T}_{-1/2,-1/2}\right)^2 + 4{\rm T}_{-1/2,1/2}{\rm T}_{1/2,-1/2}}\,\right].
\end{eqnarray}
Other thermodynamic quantities, such as the total magnetization $m$ normalized per one magnetic particle and the entropy $S$, can be immediately obtained from the Gibbs free energy~(\ref{eq:G}) by using the standard thermodynamic relations
\begin{eqnarray}
\label{eq:magnetizationEntropy}
m = -\frac{1}{4N}\left(\frac{\partial G}{\partial B}\right)_{T}, \qquad
S = -\left(\frac{\partial G}{\partial T}\right)_{B}.
\end{eqnarray}
The sublattice magnetization $m_I$ and $m_e$ normalized per one Ising spin and one mobile electron, respectively, can also be obtained as derivatives of the Gibbs free energy~(\ref{eq:G}):
\begin{eqnarray}
\label{eq:mI}
m_I \!\!\!&=&\!\!\! g_I\mu_{\rm B}\langle\sigma_k^z\rangle =
-\frac{g_I\mu_{\rm B}}{N}\left(\frac{\partial G}{\partial h_I}\right)_{T}, \\
\label{eq:me}
m_e \!\!\!&=&\!\!\! \frac{g_e\mu_{\rm B}}{3}\langle S_k^z\rangle = \frac{g_e\mu_{\rm B}}{3}\bigg\langle\frac{1}{2}\!\sum_{j\in \tilde{3}}(n_{kj,\uparrow}-n_{kj,\downarrow})\bigg\rangle =
-\frac{g_e\mu_{\rm B}}{3N}\left(\frac{\partial G}{\partial h_e}\right)_{T}.
\end{eqnarray}
In view of this notation, the total magnetization $m$ normalized per one magnetic particle of the model, appearing in Eq.~(\ref{eq:magnetizationEntropy}), can alternatively be expressed as
\begin{eqnarray}
\label{eq:mag}
m = \frac{1}{4}\left(m_I + 3m_e\right).
\end{eqnarray}
It is obvious from the set of Eqs.~(\ref{eq:TMatrixElements}), (\ref{eq:G}), (\ref{eq:mI})--(\ref{eq:mag}) that the total magnetization of the system basically depends on the Land\'e $g$-factors $g_I$ and $g_e$ of the localiezed Ising spins and mobile electrons, besides a mutual interplay between the interaction terms $J$, $t$, $U$ and the size of the external field $B$. Indeed, the total magnetization normalized per one magnetic particle of relevant ground states discussed in Sec.~\ref{sec:3} equal to
\begin{eqnarray}
m_{\rm CFM} = \frac{\mu_{\rm B}}{8}\left(g_I + 3g_e\right), \quad m_{\rm QFM} = \frac{\mu_{\rm B}}{8}\left(g_I + g_e\right), \quad m_{\rm CFRI} = \frac{\mu_{\rm B}}{8}\left(g_I - 3g_e\right), \quad m_{\rm QFRI} = \frac{\mu_{\rm B}}{8}\left(g_I - g_e\right).
\end{eqnarray}

\begin{figure}[h!]
\centering
\vspace{-0.1cm}
  \includegraphics[width=1.0\textwidth]{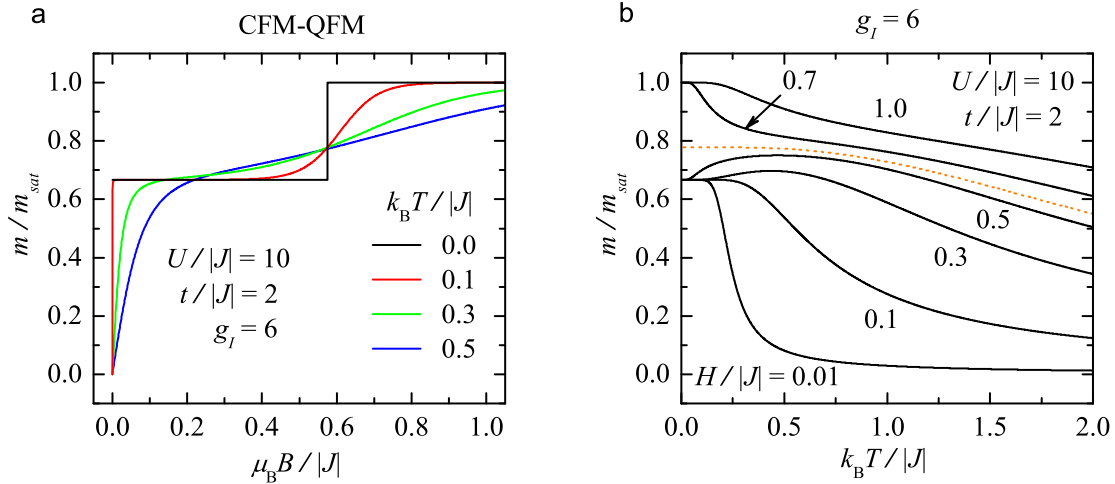}
  \vspace{-0.75cm}
\caption{The total magnetization of the model with the ferromagnetic exchange interaction $J<0$ normalized with respect to its saturation value (a) as a function of the magnetic field at a few temperatures and (b) as a function of the temperature at a few magnetic fields, by assuming the fixed values of the Coulomb term $U/|J| = 10$, the hopping term $t/|J| = 2$ and the Land\'e $g$-factor of the Ising spins $g_I = 6$. Dotted curve shown in Fig.~(b) correspond to the critical field~(\ref{eq:Bc_ferro}).}
\label{fig:3}
\end{figure}

\begin{figure}[t!]
\centering
\vspace{-0.2cm}
  \includegraphics[width=1.0\textwidth]{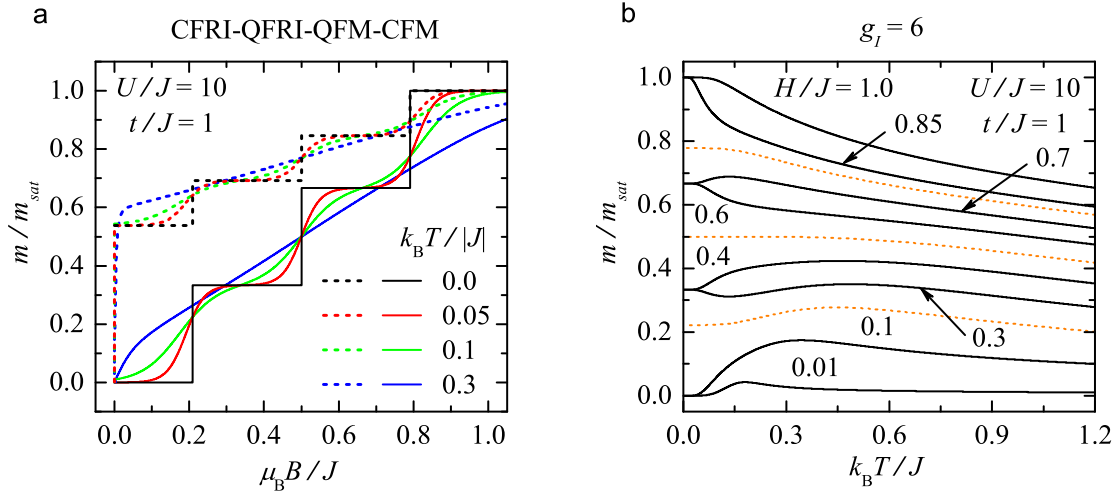}
  \vspace{-0.75cm}
\caption{The total magnetization of the model with the antiferromagnetic exchange interaction $J>0$ normalized with respect to its saturation value (a) as a function of the magnetic field at a few temperatures and (b) as a function of the temperature at a few magnetic fields, by assuming the fixed values of the Coulomb term $U/J = 10$ and the hopping term $t/J = 1$. The dotted and solid lines shown in Fig.~(a) correspond to the Land\'e $g$-factors of the Ising spins $g_I = 6$ and $g_I = 20$, while all curves plotted in Fig.~(b) correspond only to the particular case $g_I = 6$. Dotted curves shown in Fig.~(b) are plotted for critical fields~(\ref{eq:Bc_FRI1-FRI2}), (\ref{eq:Bc_UFM-FRI2}) and ~(\ref{eq:Bc_ferro}) (from bottom to top).}
\label{fig:4}
\vspace{-0.0cm}
\end{figure}

Next, let us turn our attention to a discussion of the magnetization process at zero as well as non-zero temperatures serving an evidence of the ground-state features of the investigated model. For this purpose, the total magnetization normalized with respect to its saturation value (the saturation magnetization is $m_{sat} = m_{\rm CFM}$) versus the magnetic field~$B$ is displayed in Figs.~\ref{fig:3}a and~\ref{fig:4}a for models with the ferromagnetic and antiferromagnetic exchange coupling $J$, respectively, by assuming the fixed value of the Coulomb term, two values of the hopping terms and a few temperatures. To be closer to real magnetic compounds, we consider hereafter that the mobile electrons have the fixed Land\'e $g$-factor $g_e = 2$ in contrast to the variable value of Land\'e $g$-factor for the Ising spins $g_I \geq 6$. In accordance to the ground-state analysis, a single plateau at $m/m_{sat}=(g_I + 2)/(g_I + 6)$ corresponding to the QFM ground state can be detected in magnetization curves of the model with $J<0$ (see Fig.~\ref{fig:3}a). On the other hand, the magnetization of the system with $J>0$ may display three subsequent plateaus at $m/m_{sat}= (g_I - 6)/(g_I + 6)$, $(g_I - 2)/(g_I + 6)$ and $(g_I + 2)/(g_I + 6)$ due to a possible sequence of field-induced phase transitions CFRI-QFRI-QFM-CFM (see Fig.~\ref{fig:4}a). Interestingly, the first plateau at $m/m_{sat}= (g_I - 6)/(g_I + 6)$ takes a zero or non-zero value depending on whether the Land\'e $g$-factor of the Ising spins is $g_I=6$ or $g_I>6$. The plateau at zero magnetization observed for $g_I=6$ can be attributed to the two-fold degeneracy of the CFRI phase detected at non-zero magnetic fields. In addition, the relatively complex magnetization scenario with three different plateaus found in the magnetization process for $J>0$ may be changed to a simpler magnetization curves with one plateau at at $m/m_{sat}=(g_I - 6)/(g_I + 6)$ in the limit $t/J\to 0$ or two plateaus at $m/m_{sat}=(g_I - 2)/(g_I + 6)$, $(g_I + 2)/(g_I + 6)$ when $t/J$ exceeds the zero-field boundary value explicitly given by Eq.~(\ref{eq:Bc_FRI1-FRI2}) (it is not shown in Fig.~\ref{fig:4}a, because it is clear from Fig.~\ref{fig:2}b).
It is also worth to mention that actual magnetization plateaus and magnetization jumps appear only at zero temperature. The increasing temperature gradually smooths the magnetization curves. To be complete, Figs.~\ref{fig:3}b and~\ref{fig:4}b show typical temperature variations of the total magnetization of models with $J<0$ and $J>0$, respectively, that evidence pronounced low-temperature variations of the magnetization when the field $B$ is very weak or fixed slightly below/above relevant critical fields.
\newpage

\begin{figure}[t!]
\centering
\vspace{0.0cm}
  \includegraphics[width=1.0\textwidth]{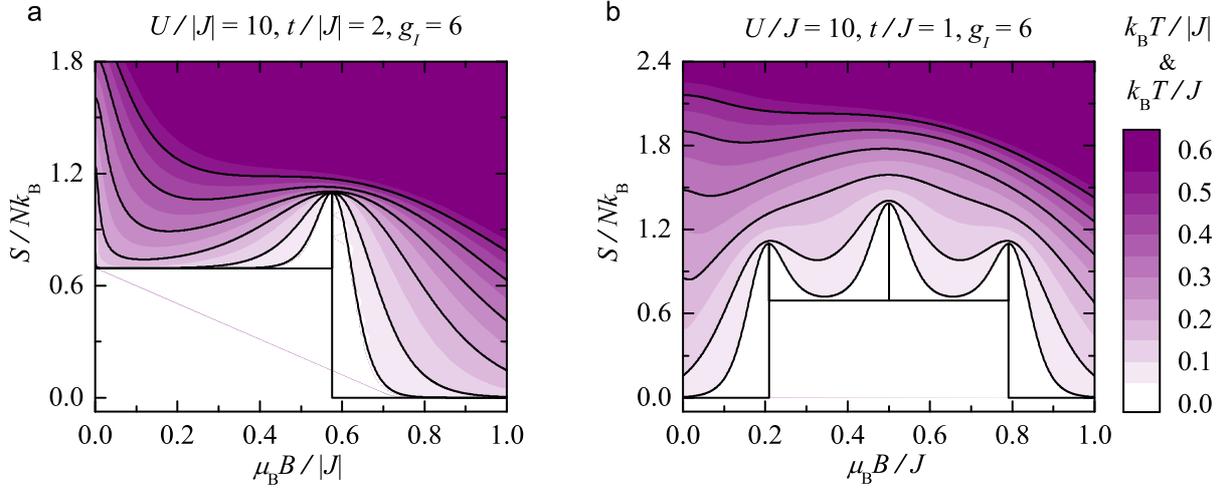}
  \vspace{-0.9cm}
\caption{Field-temperature dependencies of the entropy $S/Nk_{\rm B}$ (a)~for the model with $J<0$ by assuming the same values of parameters as in Fig.~\ref{fig:3} and (b)~for the model with $J>0$ by assuming the same values of parameters as in Fig.~\ref{fig:4}b. Solid curves correspond to isothermal changes of the entropy under the magnetic field variation at the various temperatures $k_{\rm B}T/|J| = k_{\rm B}T/J = 0, 0.05, 0.1, 0.2, 0.3, 0.4, 0.5$ (from bottom to top).}
\label{fig:5}
\end{figure}

Now, let us look to the entropy of the model at individual ground states as well as at finite temperatures. For this purpose, the entropy $S$ normalized per one Ising spin as a function of the magnetic field and temperature is displayed in Fig.~\ref{fig:5} for two different magnetization scenarios discussed previously. Solid curves plotted in this figure pick up several isothermal changes of the entropy upon varying the magnetic field. Evidently, low-temperature regions quite well reflect the field-induced phase transition QFM-CFM (Fig.~\ref{fig:5}a) and a sequence of three field-induced transitions CFRI-QFRI-QFM-CFM (Fig.~\ref{fig:5}b) detected in ground states of the ferromagnetic and antiferromagnetic version of the model, respectively. In particular, the entropy isotherms exhibit irregular dependencies with one or three pronounced peaks located around the critical fields~(\ref{eq:Bc_ferro}), (\ref{eq:Bc_UFM-FRI2}) and (\ref{eq:Bc_FRI1-FRI2}) when the temperature is low enough. At zero temperature, the continuous entropy isotherms split into the horizontal lines $S/Nk_{\rm B} = 0$, $S/Nk_{\rm B} = \ln 2\approx 0.693$ and vertical lines with the amplitudes $S/Nk_{\rm B} = \ln 3 \approx 1.099$, $S/Nk_{\rm B} =\ln 4 \approx 1.386$. In consonance with the ground-state analysis presented in Sec.~\ref{sec:3}, the zero-temperature isotherms confirm the existence of the long-range order in CFRI and CFM phases, in contrast to the horizontal lines $S/Nk_{\rm B} = \ln 2$ observed in other field regions that clearly point to the macroscopic degeneracy of the QFRI and QFM phases. Finally, the amplitude $S/Nk_{\rm B} = \ln 3$ of the vertical zero-temperature isotherms reflects the macroscopic degeneracy of the system found at phase transitions between the macroscopically ordered CFM (CFRI) phase and the macroscopically degenerate QFM (QFRI) phase, while the amplitude $S/Nk_{\rm B} = \ln 4$ reflects the degeneracy of the system at the phase transition between the macroscopically degenerated QFRI and QFM phases.

\section{Magnetocaloric properties}
\label{sec:5}

It is well known that the non-zero residual entropy may give rise to an enhanced MCE, which is accompanied by a relatively fast cooling (or heating) of the system upon the adiabatic or isentropic variation of the external magnetic field.
In general, the existence of the MCE can be demonstrated by a large isothermal entropy change or/and a large adiabatic temperature change when the system is exposed to a varying magnetic field. For the studied spin-electron chain, the isothermal entropy change ($\Delta S_T$) and the adiabatic temperature change ($\Delta T_{ad}$) upon the magnetic-field variation $\Delta B\!: 0\to B$ can be rigorously calculated by using the thermodynamic relation~(\ref{eq:magnetizationEntropy}) for the magnetic entropy:
\begin{eqnarray}
\label{eq:dS_T}
\Delta S_{T} (T, \Delta B)= S(T, B\neq 0) - S(T, B = 0),
\\
\label{eq:dT_ad}
\Delta T_{ad} (S, \Delta B)= T(S, B\neq 0) - T(S, B = 0).
\end{eqnarray}
Recall that the former relation~(\ref{eq:dS_T}) is valid if the temperature $T$ of the model is constant, while the latter one (\ref{eq:dT_ad}) satisfies the adiabatic condition $S(T, B\neq 0) = S(T, B = 0)$.

\begin{figure}[t!]
\centering
\vspace{-0.2cm}
  \includegraphics[width=1.0\textwidth]{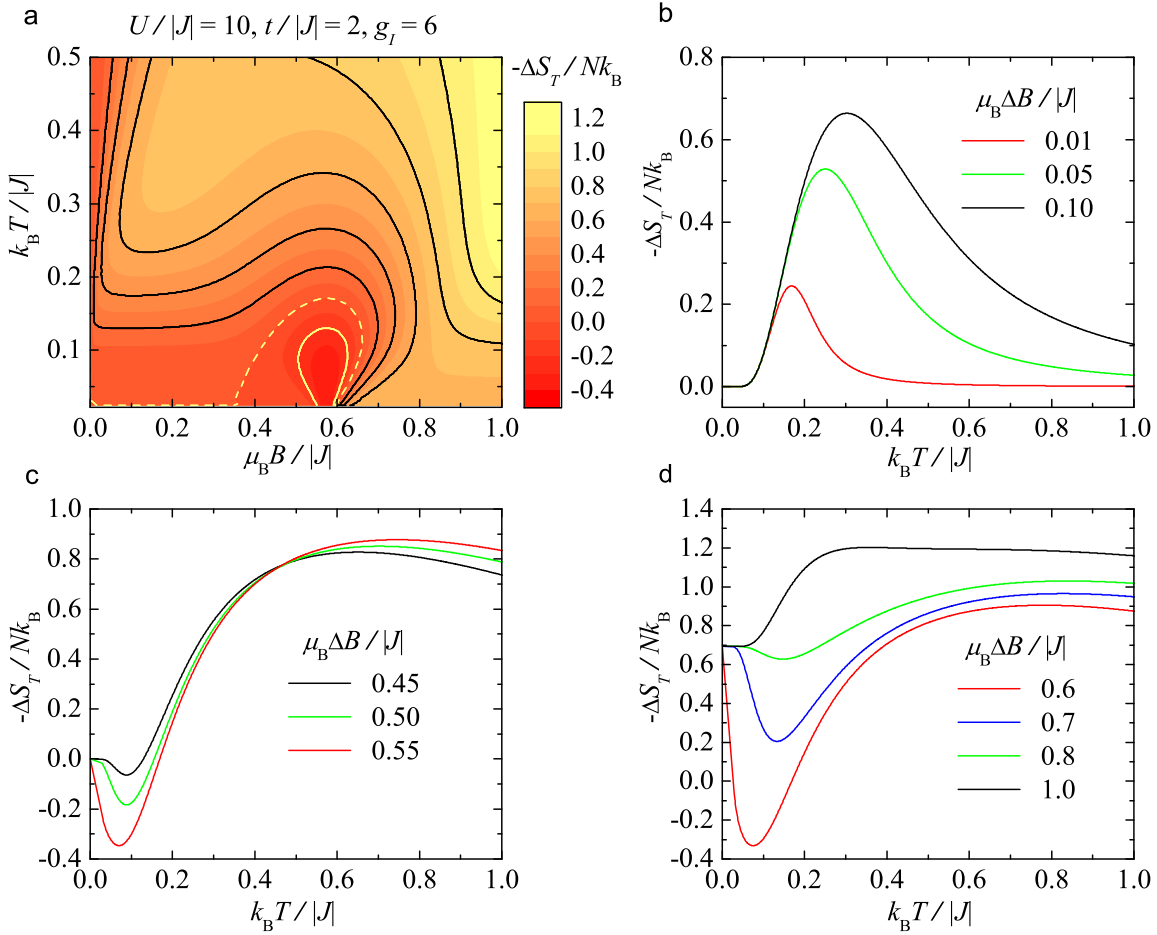}
  \vspace{-0.75cm}
\caption{The isothermal entropy change $-\Delta S_T/Nk_{\rm B}$ as a function of the magnetic field and temperature for the model~with $J<0$ and the same values of parameters as in Figs.~\ref{fig:3} and~\ref{fig:5}a. (a) A density plot of $-\Delta S_T/Nk_{\rm B}$ in the $B-T$ plane; the displayed curves correspond to the fixed values $-\Delta S_T/Nk_{\rm B} = -0.2$ (yellow solid curve), $-\Delta S_T/Nk_{\rm B} = 0$ (yellow dashed curve) and $-\Delta S_T/Nk_{\rm B} = 0.2, 0.4, 0.6, 0.8, 1.0$ (black solid curves), from bottom to top. (b)--(d) Thermal dependencies of $-\Delta S_T/Nk_{\rm B}$ for several magnetic-field changes $\Delta B\!: 0\to B$.
}
\label{fig:6}
\vspace{-0.0cm}
\end{figure}

First, let us turn our attention to the isothermal entropy changes, achieved by increasing the magnetic field from zero to finite value, that are depicted in Fig.~\ref{fig:6} for the model with the ferromagnetic Ising interaction $J<0$. To enable a direct comparison, the model parameters $t$, $U$ and $g_I$ are chosen so as to match magnetization scenario and thermal dependencies of the total magnetization plotted in Fig.~\ref{fig:3}. It is obvious from Fig.~\ref{fig:6} that the magnetocaloric potential $\Delta S_{T}$ may be negative or positive depending on values of the applied magnetic field and temperature. Thus, the investigated model exhibits both conventional ($-\Delta S_{T}>0$) and inverse ($-\Delta S_{T}<0$) MCE. More specifically, at high temperatures, $-\Delta S_{T}$ is solely positive and gradually increases with decreasing temperature and increasing magnetic field due to suppression of a spin and electron disorder by the applied field. However, as temperature falls bellow a certain value, $-\Delta S_{T}$ starts to abruptly decrease with further decreasing the temperature. Depending on the value of applied field, $-\Delta S_{T}$ versus temperature either monotonously decreases or it exhibits a local minimum before the temperature approaches the zero. The former thermal dependencies of $-\Delta S_{T}$, that reveal broad maxima in $-\Delta S_{T}(T)$ curves, can be detected at small and also high fields when they are selected far enough from the field-induced phase transition between the QFM and CFM phases (see Fig.~\ref{fig:6}b and also the curve corresponding to $\mu_{\rm B}\Delta B/|J|\!: 0\to 1$ in Fig.~\ref{fig:6}d). By contrast, latter ones can be observed near the phase boundary QFM-CFM. Moreover, the closer is the magnetic-field change to the value matching the critical field~(\ref{eq:Bc_ferro}), the more pronounced minimum can be observed in relevant thermal dependencies of $-\Delta S_{T}$ (see Figs.~\ref{fig:6}c,d). Negative amplitudes of minima in low-temperature parts of $-\Delta S_T (T)$ curves, detected in a relatively wide range of the magnetic-field changes $\mu_{\rm B}\Delta B/|J| \in (0.35, 0.67)$, clearly indicate a large inverse MCE around the field-induced phase transition between the QFM and CFM phases (see Fig.~\ref{fig:3}a). The origin of this phenomenon can be attributed to strong thermal fluctuations of mobile electrons leading to steep thermally-induced variations of total magnetization in this region (compare $-\Delta S_{T}(T)$ curves in Figs.~\ref{fig:6}c,d with corresponding thermal variations of $m/m_{sat}$ in Fig.~\ref{fig:3}b).

\begin{figure}[t!]
\centering
\vspace{-0.2cm}
  \includegraphics[width=1.0\textwidth]{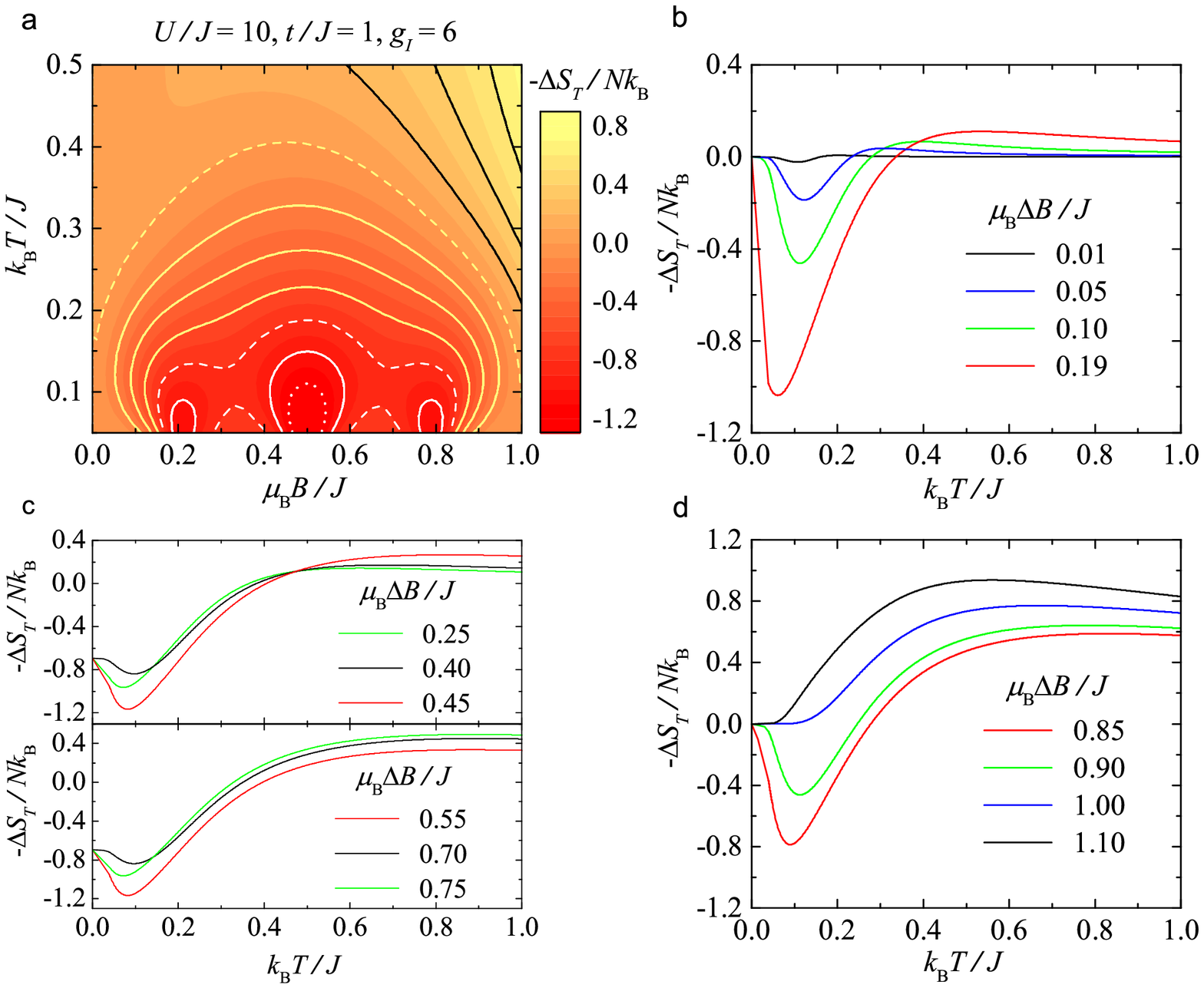}
  \vspace{-0.75cm}
\caption{The isothermal entropy change $-\Delta S_T/Nk_{\rm B}$ as a function of the magnetic field and temperature for the model~with $J>0$ and the same values of parameters as in Figs.~\ref{fig:4} and~\ref{fig:5}b. (a) A density plot of $-\Delta S_T/Nk_{\rm B}$ in the $B-T$ plane; the displayed curves correspond to the fixed values $-\Delta S_T/Nk_{\rm B} = -1.2$ (white dotted curve), $-\Delta S_T/Nk_{\rm B} = -1.0$ (white solid curve), $-\Delta S_T/Nk_{\rm B} = -0.8$ (white dashed curve), $-\Delta S_T/Nk_{\rm B} = -0.6, -0.4, -0.2$ (yellow solid curves), $-\Delta S_T/Nk_{\rm B} = 0$ (yellow dashed curve) and $-\Delta S_T/Nk_{\rm B} = 0.2, 0.4, 0.6$ (black solid curves), from bottom to top. (b)--(d) Thermal dependencies of $-\Delta S_T/Nk_{\rm B}$ for several magnetic-field changes $\Delta B\!: 0\to B$.
}
\label{fig:7}
\end{figure}

Both conventional and inverse MCE can also be observed in the model with the antiferromagnetic Ising interaction $J>0$. For illustration, we present in Fig.~\ref{fig:7} isothermal entropy changes versus magnetic field and temperature for to the same model parameters as in Fig.~\ref{fig:4}. It is obvious from this figure that the inverse MCE can be found within some range of low temperatures along the whole magnetization process up to a certain magnetic field, above which it vanishes and just the conventional MCE at moderate temperatures can be observed. As expected, the inverse MCE is most pronounced in a close vicinity of field-induced phase transitions between relevant phases due to strong vigorous thermal excitations of magnetic particles from a ground state to a first-excited state. The conventional MCE gradually increases with the increasing magnetic field as the Zeeman's energy suppresses thermal fluctuations of magnetic particles in the system.

\begin{figure}[t!]
\centering
\vspace{-0.2cm}
  \includegraphics[width=1.0\textwidth]{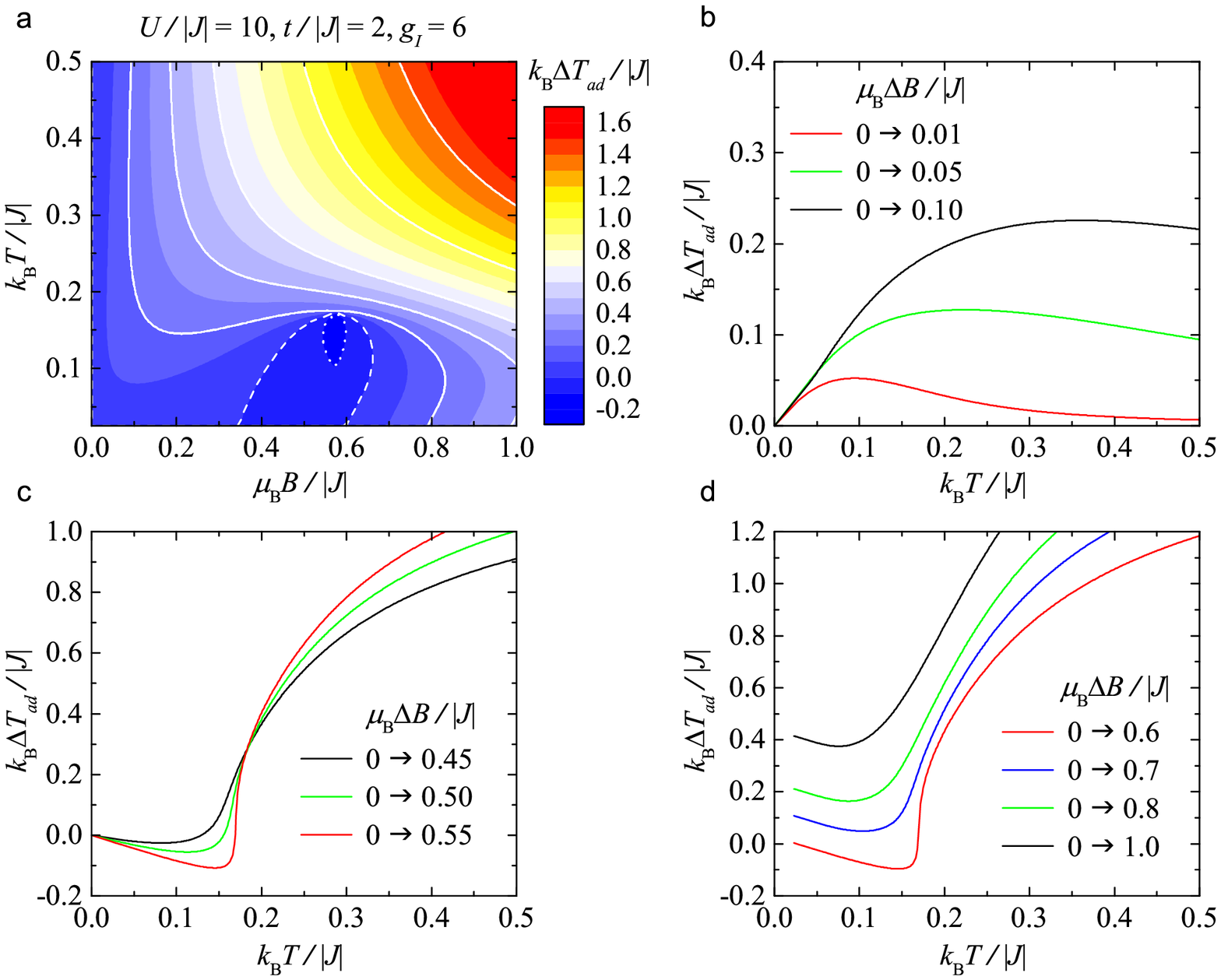}
  \vspace{-0.25cm}
\caption{The adiabatic temperature change $k_{\rm B}\Delta T_{ad}/J$ as a function of the magnetic field and temperature for the model~with $J<0$ and the same values of model parameters as in Fig.~\ref{fig:6}. (a) A density plot of $k_{\rm B}\Delta T_{ad}/|J|$ in the $B-T$ plane; white curves correspond to the fixed values $k_{\rm B}\Delta T_{ad}/|J| = -0.1$ (dotted curve), $k_{\rm B}\Delta T_{ad}/|J| = 0$ (dashed curve) and $k_{\rm B}\Delta T_{ad}/|J| = 0.2, 0.4, 0.6, 1.0, 1.4$ (solid curves), from bottom to top. (b)--(d)~Thermal dependencies of $k_{\rm B}\Delta T_{ad}/|J|$ for the same magnetic-field changes $\Delta B\!: 0\to B$ as in Fig.~\ref{fig:6}.
}
\label{fig:8}
\vspace{-0.0cm}
\end{figure}

\begin{figure}[t!]
\centering
\vspace{-0.2cm}
  \includegraphics[width=1.0\textwidth]{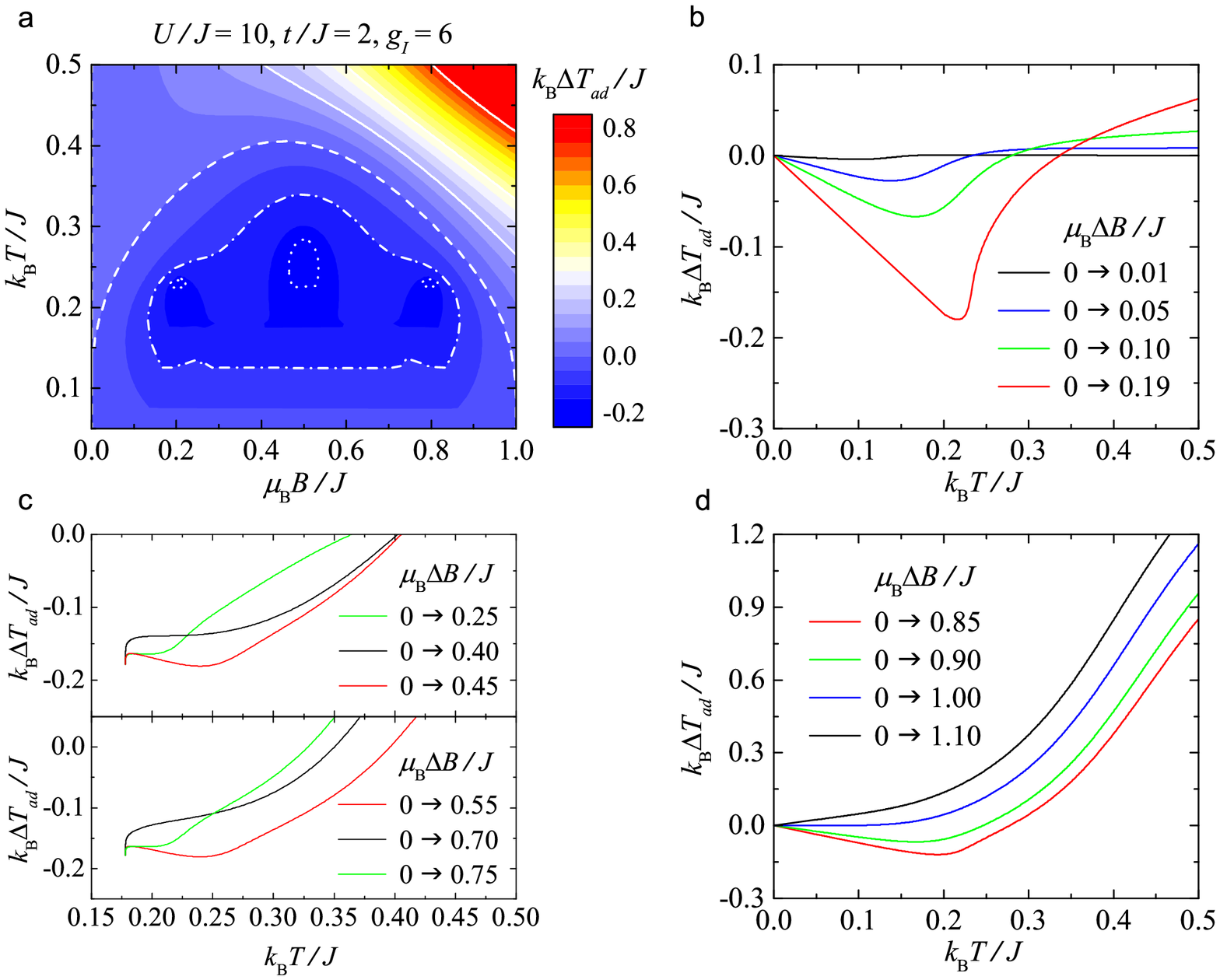}
  \vspace{-0.75cm}
\caption{The adiabatic temperature change $k_{\rm B}\Delta T_{ad}/J$ as a function of the magnetic field and temperature for the model~with $J>0$ and the same values of model parameters as in Fig.~\ref{fig:7}. (a) A density plot of $k_{\rm B}\Delta T_{ad}/J$ in the $B-T$ plane; white curves correspond to the fixed values $k_{\rm B}\Delta T_{ad}/J = -0.2$ (dotted curves), $k_{\rm B}\Delta T_{ad}/J = -0.1$ (dashed-dotted curve), $k_{\rm B}\Delta T_{ad}/J = 0$ (dashed curve) and $k_{\rm B}\Delta T_{ad}/J = 0.2, 0.4, 0.8$ (solid curves), from bottom to top. (b)--(d)~Thermal dependencies of $k_{\rm B}\Delta T_{ad}/J$ for the same magnetic-field changes $\Delta B\!: 0\to B$ as in Fig.~\ref{fig:7}.
}
\label{fig:9}
\end{figure}

To discuss the MCE, one may alternatively investigate the adiabatic temperature change of the system at various magnetic-field changes $\Delta B\!:0 \to B$. The low-temperature and low-field variations of this magnetocaloric potential for both the models with $J<0$ as well as $J>0$ are shown in Figs.~\ref{fig:8} and~\ref{fig:9}, respectively. Note that all curves plotted in these figures were calculated using Eq.~(\ref{eq:dT_ad}) by keeping the entropy constant. In accordance to the previous discussion, the system with $J<0$ displays an enhanced inverse MCE in a range of the magnetic fields $\mu_{\rm B}B/|J|\in (0.35, 0.67)$ for the temperatures $k_{\rm B}T/|J|< 0.18$ and a conventional MCE for other fields and temperatures. Indeed, the system cools down most significantly upon the application of relatively small magnetic field $\mu_{\rm B}B/|J|\approx 0.576$.
For this particular field, the adiabatic temperature change achieves the value $k_{\rm B}\Delta T_{ad}/|J|\approx -0.161$. On the other hand, the antiferromagnetic counterpart of system (i.e., the system with $J>0$) cools down along whole magnetization process up to the field $\mu_{\rm B}B/|J|\approx 1$, above which just positive adiabatic temperature changes can be detected.  As predicted previous discussion, the most pronounced negative adiabatic temperature changes occur in a vicinity of the field-induced phase transitions between relevant phases. The largest negative adiabatic temperature change $k_{\rm B}\Delta T_{ad}/|J|\approx -0.245$ can be found approximately for the magnetic-field change $\mu_{\rm B}\Delta B/|J|\!:0 \to 0.491$, i.e., just below the phase transition between the QFRI and QFM ground states. In addition, for both particular cases $J<0$ and $J>0$, $\Delta T_{ad}$ versus temperature plots end at zero value in the asymptotic limit of zero temperature whenever the entropy of the ground state observed at non-zero field is the same as the entropy of the zero-field ground state (compare the results in Figs.~\ref{fig:8}b, c and~\ref{fig:9}b, d with the corresponding zero-temperature entropy curves shown in Fig.~\ref{fig:5}). Otherwise, the adiabatic temperature changes end at different values $k_{\rm B}T_{ad}/|J| >0$ (if $J<0$) and at the fixed negative value $k_{\rm B}T_{ad}/J = -0.178$ (if $J>0$) at the temperatures $k_{\rm B}T/|J| = 0.023$ and $k_{\rm B}T/J = 0.178$, respectively (see Figs.~\ref{fig:8}d and~\ref{fig:9}c). In these particular cases, the magnetocaloric potential $\Delta T_{ad}$ cannot be defined below
aforementioned temperatures, because there is no temperature starting or end point in the adiabatic process. This intriguing behaviour is evidently caused by residual entropy $S/Nk_{\rm B} = \ln 2$ detected within macroscopically degenerate QFM and QFRI ground states (see Fig.~\ref{fig:5}).

\section{Concluding remarks}
\label{sec:6}

The present paper deals with the ground-state and magnetocaloric properties of the double-tetrahedral chain, where the nodal lattice sites occupied by the localized Ising spins regularly alternate with three equivalent lattice sites available for three mobile electrons. By using the standard transfer-matrix method, we have analytically derived exact results for the basic thermodynamic quantities of the system, namely for the Gibbs free energy, the total and sublattice magnetization as well as the entropy. It has been shown that the ground-state phase diagram of the model consists of two ferromagnetic (CFM and QFM) and two ferrimagnetic (CFRI and QFRI) ground phases. Two of them, namely QFM and QFRI phases, are macroscopically degenerate due to chiral degrees of freedom of the mobile electrons. In accordance to the ground-state analysis, several plateaus have been observed in magnetization process of the model, depending on the mutual interplay between model parameters and the applied longitudinal magnetic field. At the critical fields corresponding to the field-induced phase transitions between relevant phases, where the magnetization jumps between relevant plateaus are present, the model exhibits a relatively high macroscopic degeneracy, which leads to the non-zero residual entropy. By using the exact solution for the magnetic entropy, we have obtained numeric results for the adiabatic entropy change and the isothermal temperature change of the system. Both magnetocaloric quantities enabled us to clarify the magnetic refrigeration efficiency of the model in a vicinity of the first-order phase transitions. The obtained results for both magnetocaloric potentials clearly indicate on the fast heating/cooling of investigated spin-electron chain during the adiabatic demagnetization/magnetization process (on a presence of the enhanced inverse MCE) in these regions due to strong thermal fluctuations of magnetic particles leading to steep thermally-induced variations of the total magnetization.

\section*{Acknowledgments}
L.G. acknowledges the financial support provided by a grant of the Slovak Research and Development Agency under the contract No. APVV-0097-12 and by a scientific grant from The Ministry of Education, Science, Research and Sport of the Slovak Republic under the contract VEGA 1/0043/16.

\end{document}